\documentstyle[prb,aps,multicol,epsfig]{revtex}

\begin{document}

\title{Different definitions of the chemical potential in the action with identical partition 
function in QCD on a lattice}

\author{Fabrizio Palumbo~\thanks{This work has been partially 
  supported by EEC under TMR contract ERB FMRX-CT96-0045}}
\address{INFN -- Laboratori Nazionali di Frascati - P.~O.~Box 13, I-00044 Frascati, ITALIA}

\date{\today}

\maketitle

\begin{abstract}

It is shown that starting from one and the same transfer matrix formulation of QCD on a lattice, it is
possible to obtain both the action of Hasenfratz and Karsch as well as an action where the
chemical potential is not coupled to the temporal links.

\vspace{0.3cm}


\end{abstract}


\begin{multicols}{2}


Numerical simulations of QCD at finite baryon density have met with long standing
difficulties~\cite{Kogu}. It is then natural to inquire whether the situation cannot be improved by
modifying the action used in the path integral. Special attention deserves the coupling of
the chemical potential, whose form is due to Hasenfratz and Karsch~\cite{Hase}. They 
noticed that if in the euclidean path integral of relativistic gauge theories regularized on a lattice
the chemical potential is coupled to the fermi fields  as in classical systems, specific counterterms are
necessary which would be extremely  unconvenient in numerical simulations. To avoid this shortcoming,
guided by the analogy between  chemical potential and  temporal gauge field, they coupled 
the chemical potential in the  same way as the gauge fields in the Wilson
regularization. They showed that this definition does not require any counterterms, at least in the free
theory. After some further analysis~\cite{Bili}, this definition has been accepted as
essentially unique for practical purposes~\cite{Mont}.

The feature of the Hasenfratz-Karsh prescription which attracts our attention is the point splitting 
which implies that the chemical potential must be coupled also to the temporal gauge links. It
has a topological consequence, in the sense that only the temporal Polyakov loops give contributions
to the path integral which depend on the chemical potential. This looks like a mere lattice artifact,
because it does not have any known counterpart in the continuum. But if the interplay between chemical
potential and Polyakov loop is an artifact it could be eliminated. This appears an interesting
possibility from the point of view of numerical simulations because a term of toplogical nature is
expected  to affect them significantly.

Therefore we started a systematic investigation of QCD at finite baryon density. We first
avoided the use of the chemical potential by considering QCD in a given baryon sector~\cite{Palu}. The
resulting effective action differs by the unconstrained action by terms with a quite different structure
with respect to the coupling of the chemical potential in the Hasenfratz-Karsch action. Encouraged by
this result we reconsidered the definition of the chemical potential starting from the transfer matrix
formalism. The passage to the path integral, as far as the fermion fields are concerned, is  based on the
insertion of the identity expressed in terms of coherent states in the trace of the transfer matrix.
But this can be done in different ways. A possibility which leads exactly to the
Hasenfratz-Karsch prescription has been fully investigated~\cite{Palu,Mitr}. The point
splitting and the exponentiation of the chemical potential are in this case a consequence of the
properties of Grassmannian kernels of fermionic operators~\cite{Creu}. But among other possibilities we
show here that there is one where there is no point splitting and the chemical potential is not coupled
to the temporal links. For Kogut-Susskind fermions in the flavour basis it is not coupled to the gauge
fields at all.  

We start from the definiton of the grand canonical partition function at finite baryon
density according to the ordered product
\begin{equation}
Z = Tr \left\{  \exp \left(  {\mu a_0 \over T} \, \hat{Q}_B \right) 
\prod_{n_0}\hat{{\cal T}}_q(n_0) \right\}.  \label{trace} 
 \end{equation}
 $a_0$ is the lattice spacing in the temporal direction, $n_0$ the temporal component of the site
position vector $n$, $N_0$ is the number of links in the temporal direction, $T$ the temperature
\begin{equation}
T= { 1 \over a_0 N_0},
\end{equation}
$ \hat{Q}_B $ is the baryon charge operator and $ \hat{{\cal T}}_q $ is the quark transfer matrix.
The pure gauge part of the transfer matrix has been omitted because it does not play any role in the
problem.
 The chemical potential $\mu$
is determined by the condition that the expectation value of the baryon number be
\begin{equation}
 q_B = Z^{-1} Tr  \left\{ \hat{Q}_B\exp (  \mu N_0 \, \hat{Q}_B )\prod_{n_0}
\hat{{\cal T}}_q(n_0) \right\}.
\end{equation}
 $ \hat{{\cal T}}_q $ can be written~\cite{Lusc,Palu} in terms of an 
auxiliary operator $\hat{T}_q(n_0) $
\begin{equation} 
\hat{{\cal T}}_q(n_0) =  {\cal J} ^{-1} \hat{T}_q^{\dagger}(n_0) \,
 \hat{T}_q(n_0+s), \label{transfer}
 \end{equation}
where ${\cal J} $ is a function of the gauge fields which will be defined later and 
$s=\pm 1$ for Wilson/Kogut-Susskind fermions respectively. This difference in sign 
does not reflect any intrinsic difference, but is only due to the different conventions adopted
by L\"uscher~\cite{Lusc} and Montvay-M\"unster~\cite{Mont}, which we maintain for easy reference. The
case of Kogut-Susskind fermions will be treated in the flavour basis, because we do not know a
suitable formulation of the transfer matrix in the spin diagonal basis. Obviously it is very simple to
reexpress the result into this basis, but this introduces awkward nonminimal gauge couplings of the type
discussed in~\cite{Palu}. 

 $\hat{T}_q $ is defined in terms of quark-antiquark creation-annihilation operators  
$\hat{x}^{\dagger}, \hat{y}^{\dagger},\hat{x},\hat{y} $ acting in a Fock space. It depends 
on the time coordinate
$n_0$ only through the dependence on it of the gauge fields. In fact the creation and
annihilation  operators do not depend on $n_0$. They depend on the spatial
coordinates ${\bf n}$ of the sites or, in the case of Kogut-Susskind in the flavour basis,
 of the blocks, and on
Dirac, flavour and color indices, $\alpha, f,c$ sometimes comprehensively represented by $I$.
 
 In the transfer matrix formalism often one has to do with quantities at
a given (euclidean) time  $n_0$. For this reason we adopt a summation convention
over spatial coordinates and intrinsic indices at fixed time. So for instance we
will write
\begin{equation}
 \hat{x}^{\dagger} M(n_0) \, \hat{x}
= \sum_{{\bf m},{\bf n},I,J} \hat{x}^{\dagger}_{{\bf m},I}
M_{{\bf m},I;{\bf n},J}(n_0) \hat{x}_{{\bf n},J},
\end{equation}
where $M$ is an arbitrary matrix.
In this notation the baryonic charge $ \hat{Q}_B $ and the 
auxiliary operator $\hat{T}_q(n_0)  $ can be written
\begin{equation} 
\hat{Q}_B = \hat{x}^{\dagger} \hat{x} - \hat{y}^{\dagger} \hat{y} \label{baryon}
\end{equation}
\begin{equation}
\hat{T}_q(n_0)  = 
\exp\left[ -\hat{x}^{\dagger} M(n_0) \, \hat{x}  
-\hat{y}^{\dagger} M(n_0) \, \hat{y} \right]
 \exp\left[ \hat{y} \, N(n_0) \,\hat{x} \right].   \label{tq}
\end{equation}
Their form is the same for Wilson and Kogut-Susskind fermions
but the matrices  $M$ and $N$ are different in the two 
cases and will be specified later. The expression of $\hat{T}_q(n_0)$ is valid in 
the gauge $U_0 = 1\!\!1$ which is not admissible. But Menotti and Pelissetto~\cite{Meno}
extended the proof of reflection positivity to the gauge $U_0 \sim 1\!\!1$ where $U_0=1\!\!1$ 
with the exception of a single time slice. In the
construction of the path integral formulation of QCD at finite baryon density we do not need to fix the
gauge, but to lighten the formalism we will nevertheless put $U_0= 1\!\!1$ and we will reinstate $U_0$ 
in the final result.
The reader can check that keeping $U_0$ in the intermediate steps one arrives at the same result,
provided some care is exercised: for instance when $U_0 \neq 1\!\!1$ the expression $  
\hat{x}^{\dagger} M(n_0) \, \hat{x}  + \hat{y}^{\dagger} M(n_0) \, \hat{y} $ appearing in 
Eq.~\ref{tq} changes and does not commute with $\hat{Q}_B$ any longer.
We anticipate that $N$ is hermitean and also $M$ is hermitean in the gauge $U_0 = 1\!\!1$. 

Before transforming the trace into a Berezin integral it is convenient to ``distribute" the
exponential of the charge operator among the factors in the trace 
\begin{equation}
Z = Tr \left\{   \prod_{n_0} \left[\hat{{\cal T}}_q(n_0)
 \exp \left( \mu \, \hat{Q}_B \right) \right]\right\}.  \label{trace1} 
 \end{equation}
Now we introduce between the factors in the above equation the identity
\begin{equation}
1\!\!1 = \int [dx^+ dx \, dy^+ dy] \exp(-x^+ x - y^+ y) |x \, y><x \, y|,
\end{equation}
where the basis vectors
\begin{equation}
|x \, y> = |\exp( - x \,\hat{x}^{\dagger}  - y \, \hat{y}^{\dagger}) >
\end{equation}
are coherent states and the $x^+,x , y^+ ,y $ are Grassmann variables. They will be labeled by the time
slice where the unit operator is introduced. For the other indices they are subject to the 
same convention as the creation and annihilation operators.

Different realizations of the path integral arise depending on the way this insertion is performed.
The one leading to the Hasenfratz-Karsch prescription has been discussed in detail~\cite{Palu,Mitr}.
Here we consider another possibility based on the following way of writing the exponential of
the charge operator 
\begin{eqnarray}
\exp(\mu a_0 \hat{Q}) &=& \int [dx^+ dx \, dy^+ dy] \exp(\delta S)
\nonumber\\
 & \cdot & \exp(-x^+ x - y^+ y)|x \, y><x \,y|
\end{eqnarray}
where
\begin{eqnarray}
\delta S & = & \left( 1- \cosh( \mu a_0)\right)(x^+ x + y^+ y) 
\nonumber\\
& + &  \sinh( \mu a_0)(x^+ x - y^+ y).  \label{schem}
\end{eqnarray}
The above expression is obtained by expanding the exponential of the charge operator, ordering
all the annihilation operators to the left of the creation ones and inserting in each monomial 
the unity between the set of annihilation and the set of creation operators. For the rightmost 
exponential of the baryon charge one has to  move the creation
operators to the leftmost position before taking the trace. Replacing the operators  by their
Grassmannian eigenvalues one gets the above result.

Now the construction of the path integral proceeds in the standard way~\cite{Lusc}, yielding the action
for zero chemical potential plus the term generated by $\delta S $. To write down this
term we must distinguish the case of Wilson fermions from that of Kogut-Susskind.

 For Wilson fermions we set $s=1$ in Eq.~\ref{transfer} and assume
\begin{eqnarray} 
M_W(n_0) & = & - \ln \left( \left( 2K \right)^{{1\over 2}} B^{-{1\over 2}}(n_0) 
\right),
\nonumber\\
N_W(n_0) &= &2 K B(n_0)^{-{1\over 2}} c(n_0) B(n_0)^{-{1\over 2}},
\end{eqnarray}
where K is the hopping parameter and
\begin{eqnarray}
B(n_0) &=& 1\!\!1  - K\sum_{j=1}^3 \left(   \hat{U}_j(n_0) T^{(+)}_j + T^{(-)}_j \hat{U}_j^+(n_0)\right)
\nonumber\\
c(n_0) &=& { 1\over 2} \sum_{j=1}^3 i\,\sigma_j \left( \hat{U}_j(n_0) T^{(+)}_j 
- T^{(-)}_j \hat{U}_j^+(n_0) \right).
\end{eqnarray}

The link operators $\hat{U}_{\mu}(n_0)$ have the standard Wilson
variables $ U_{\mu}(n) $ as spatial matrix elements
\begin{equation}
\left(\hat{U}_{\mu}(m_0) \right)_{{\bf m},{\bf n}} = \delta_{{\bf m},{\bf n}} U_{\mu}(m).
\end{equation}
We have introduced the translation operators $T_j^{(\pm)}$  with matrix elements
\begin{equation}
\left( T_j^{(\pm)} \right)_{{\bf n}_1,{\bf n}_2} =  
\delta_{{\bf n}_2, {\bf n}_1 \pm e_j},  \label{transl}
\end{equation}
 $e_{\mu}$ being the unit vectors 
\begin{equation}
\left (e_{\mu}\right)_{\nu} = \delta_{\mu,\nu}.
\end{equation}
Next we define the quark field $q$ by the transformation
\begin{eqnarray}
x &=& B^{{1\over 2}} P^{(+)}_0 q, \,\,\,y = B^{{1\over 2}} P^{(-)}_0 q
\nonumber\\
\overline{q} &=& q^{\dagger} \gamma_0 \label{transf}
\end{eqnarray}
where
\begin{equation}
P_0^{(\pm)} = { 1\over 2} \left( 1 \pm \gamma_0 \right).
\end{equation}
The jacobian of this transformation is the function ${\cal J}$ introduced in Eq.~\ref{transfer}.
The partition function takes the form
\begin{equation}
Z_W= \int [d \overline{q} d q] \exp \left( S_W + \delta S_W \right),
\end{equation}
where
\begin{eqnarray}
S_W &=& \sum_n  \Big\{ K \sum_{\mu} \left[
\overline{q}(n)(1+\gamma_{\mu})U_{\mu}(n) q(n + e_{\mu}) \right. 
\nonumber\\ 
& &  \left. + \overline{q}(n + e_{\mu})(1-\gamma_{\mu}) U_{\mu}^{\dagger}(n)q(n) \right] 
- \overline{q}(n) q (n)  \Big\}
\end{eqnarray}
is the Wilson action with Wilson parameter $r=1$ and zero chemical potential and the contribution of the
chemical potential is obtained from Eq.~\ref{schem} 
\begin{eqnarray}
 \delta S_W & = & \sum_{n_0} \big\{ \overline{q}(n_0) \big[ \left( 1- \cosh (\mu a_0)\right) 
\nonumber\\
 &  + & \sinh (\mu a_0) \gamma_0 \big] B(n_0) q(n_0) \big\}. 
\end{eqnarray}
The first term breaks the chiral symmetry, but this is only a consequence of the breaking by the 
Wilson term. Indeed we will see that the corresponding term for Kogut-Susskind fermions respects
the chiral invariance. Notice the
"plus" sign in the exponential of the action, to comply with L\"uscher's convention.

The case of Kogut-Susskind fermions will studied only in the flavour
basis. The $n_{\mu}$ are now the block coordinates and the gauge fields are defined on the block links.
We set $s=-1$ in Eq.~\ref{transfer} and assume~\cite{Palu}
\begin{eqnarray} 
& & M_{KS}(n_0) = 0
\nonumber\\
& & N_{KS}(n_0)=  \Big\{ \sum_{j=1}^3 \left[  \gamma_5 \otimes t_5 t_j  
+ \gamma_j \left(  P^{(-)}_j \hat{U}_j(n_0)T^{(+)}_j \right. \right. 
\nonumber\\
& & \,\,\, \left. \left. - P^{(+)}_j T^{(-)}_j \hat{U}_j^+(n_0)  
\right)\right] 
 +  { m \over K} \, 1\!\!1 \otimes 1\!\!1 +  \gamma_5 \otimes t_5 t_0  \Big\},
\end{eqnarray}
where $m$ is the quark mass parameter and  K the hopping parameter.
In the tensor product the $\gamma$-matrices  act on Dirac indices, while the $t$-matrices  
\begin{equation}
t_{\mu}= \gamma_{\mu}^T
\end{equation}
act on flavor indices. The projection operators $ P_{\mu}^{(\pm)}$ are given by
 \begin{equation}
P_{\mu}^{(\pm)}  ={ 1\over 2} \left[ 1\!\!1 \otimes 1\!\!1 \pm
\gamma_{\mu} \gamma_5 \otimes  t_5t_{\mu} \right].
\end{equation}
The quark field $q$ is obtained by the transformation
\begin{eqnarray}
x &=& 4 \sqrt{K} P_0^{(+)} q, \,\,\,y^+ = 4 \sqrt{K} P_0^{(-)} q
\nonumber\\
\overline{q} &=& q^{\dagger} \gamma_0 
\end{eqnarray}
whose jacobian is the function ${\cal J}$ introduced in Eq.~\ref{transfer}.
The partition function takes the form
\begin{equation}
Z_{KS}= \int [d \overline{q} d q] \exp \left[- 16 \left(\,  S_{KS} + \delta S_{KS} \right) \right],
\end{equation}
where
 \begin{eqnarray}
S_{KS} & =&{1 \over 2} K \Big\{\sum_n {\overline q}(n) \left( \gamma_{\mu} \otimes 1\!\!1 - \gamma_5 
\otimes t_5 t_{\mu} \right)  U_{\mu}(n) q(n  + e_{\mu})
\nonumber\\
& - &  {\overline q}(n )  \left( \gamma_{\mu} \otimes 1\!\!1 + \gamma_5 
\otimes t_5  t_{\mu} \right) U_{\mu}^{\dagger}(n - e_{\mu}) q(n - e_{\mu} )
\nonumber\\
& + &  2 \, {\overline q}(n ) \gamma_5  \otimes t_5 t_{\mu} \, q(n ) +  a m \,
 {\overline q}(n ) 1\!\!1  \otimes 1\!\!1 q(n ) \Big\}
\end{eqnarray}
is the Kogut-Susskind action with zero chemical potential
and the chemical potential contribution is obtained from Eq.~\ref{schem}
\begin{eqnarray}
\delta S_{KS} & = & - K \Big\{\sum_n {\overline q}(n )\left[ \, \left( 1 - \cosh (\mu a_0)  \right)
\gamma_5
\otimes t_5 t_0   \right.
\nonumber\\
& & \left. + \sinh (\mu a_0) \gamma_0 \otimes 1\!\!1   \, \right] q(n)  \Big\}.
\end{eqnarray}
The factor $16$ in front of the action accounts for the fact that the volume element
with Kogut-Susskind fermions is 16 times larger than in the Wilson case.

 Let us emphasize that the partition functions with the present action and with that of Hasenfratz  and
Karsch are identical with one another. It is worth while checking this identity in the free case and
 comparing to the ``naive" definition.
 For simplicity we will consider only the Wilson case neglecting the mass and the spatial Wilson
term. But notice that we cannot omit also the temporal one, because otherwise we cannot construct the
transfer matrix. We then consider the quark matrix
\begin{equation}
Q = \gamma_0 ( a_0 \nabla_0 + \sigma ) + a_0 \gamma \cdot \nabla 
+ { 1\over 2} a_0^2 \Box_0 + 1- R
\end{equation}
where
\begin{eqnarray}
(\nabla_{\mu} f)(n) &=& { 1\over 2 a_{\mu}} \left[ f(n+e_{\mu}) - f(n-e_{\mu} ) \right]
\nonumber\\
 (\Box_{\mu} f)(n) &=&{ 1\over  a_{\mu}} \left[ f(n+e_{\mu}) + f(n-e_{\mu} ) - 2 \right]
\end{eqnarray}
and
\begin{eqnarray}
\sigma &= &\sinh(\mu a_0), \,\,\, R = \cosh(\mu a_0), \,\,\,{\rm new \, definition}
\nonumber\\
\sigma &= &\mu a_0, \,\,\, R = 1,  \,\,\,{\rm ``naive" definition}.
\end{eqnarray}
We now evaluate the  energy density ${\cal E}$ at zero temperature, setting $a_{\mu}=a$. Normalizing at
zero baryon density 
\begin{equation}
{\cal E}=- { 1\over 2 \pi^3 a^4} \int_{-\pi}^{\pi} d^3 q \, s^2 \left[{\cal I}(\mu )- {\cal I}(0)
\right],
\end{equation}
where
\begin{equation}
s^2 = \sum_j \left( 1- \cos q_j  \right),
\end{equation}
\begin{eqnarray}
{\cal I}(\mu ) &=& { 1 \over 2 \pi} \int_{-\pi}^{\pi} dq_0 \left[( \sin q_0 - i \sigma)^2 + s^2
+ (\cos q_0 - R)^2 \right]^{-1}
\nonumber\\
& =& { 1\over 2 \sqrt{ A^2 - B^2}}[ 1- \theta(R-A)].
\end{eqnarray}
$\theta$ is the step function.
With the new definition
\begin{eqnarray}
R-A &=&{ 1\over 2} \left[ s^2+2(1-\cosh(\mu a)\right] \sim { 1\over 2} \left[ s^2  -(\mu a)^2\right]
\nonumber\\
A^2-B^2 &=& s^2 \left[1+{ 1\over 4}s^2\right]
\end{eqnarray}
while with the `naive" definition
\begin{eqnarray}
R-A &=&{ 1\over 2} \left[ s^2  -(\mu a)^2 \right]
\nonumber\\
A^2-B^2 &=& s^2 \left[1+{ 1\over 4}s^2 - { 1\over 2} (\mu a)^2 + { (\mu a)^4\over 4 s^2}\right].
\end{eqnarray}
Therefore in the ``naive" case the energy density has a quadratic divergence, while with the new
definition 
\begin{equation}
{\cal E}= { 2 \over \pi^2} \mu^4
\end{equation}
is equal to the value obtained by the Hasenfratz-Karsch prescription in the presence of 
the temporal Wilson term (without the spatial one).

In conclusion we have shown how starting from one and the same transfer matrix we can realize
identical path integrals with different actions for QCD at finite baryon density. In the action
constructed in the present work the chemical potential is not coupled  to the temporal
links, in contrast to the Hasenfratz-Karsh prescription.  
 It might therefore  have quite different properties in numerical simulations which it could
be interesting to investigate.

\end{multicols}

\end{document}